\documentstyle[aps,pre]{revtex}
\begin{document}
\title{Nonequilibrium corrections in the pressure tensor due to an
energy flux} 
\author{Raquel Dom\'{\i}nguez-Cascante \thanks{E-mail:
raquel@ulises.uab.es} and Jordi Faraudo \thanks{E-mail: jordi@ulises.uab.es}}
\address{Departament de F\'{\i}sica\\
Universitat Aut\`onoma de Barcelona\\
08193 Bellaterra, Spain}
\maketitle
\begin{abstract}
The form ${\bf P}= a(u,v) {\bf U} + b(u,v) \vec{J} \vec{J}$ for the
pressure tensor for a system submitted to an energy flux $\vec{J}$,
widely used for anisotropic radiation and
proposed to be more general in \cite{QK} has been recently questioned
by Nettleton \cite{Critica}. We provide a physical basis, in a
completely different way, for this expression for anisotropic radiation and
ultrarelativistic gases and we criticize some previous physical
interpretations. We recall the necessity of an understanding of this
kind of expressions in a thermodynamic framework.
\end{abstract}
\pacs{05.70.Ln, 05.20.Gg, 44.10.+i}

\section*{}
Radiation hydrodynamics \cite{Pomraning} is a subject of great interest in
astrophysics, cosmology and plasma physics. However, the numerical
methods proposed to solve the transfer equation for the specific
radiation intensity $\cal{I}$ are in many cases computationally too
expensive. Therefore, one usually considers the equations for the moments of
$\cal{I}$ up to a given order $m$ \cite{Lever,Anile}. Due to the
dependence of the equation for the moment $m$ on the moment $m+1$ one needs to
introduce a closure relation. If only the energy density $u$ ($m=0$) and
the energy density flux $\vec{J}$ ($m=1$) are considered, one must introduce a
closure relation for the pressure tensor ${\bf P}$ ($m=2$). The usual
procedure is to introduce the so-called Eddington factor $\chi$ defined by:
\begin{equation}
{\bf P} = u \left [\frac{1-\chi}{2} {\bf U} +
\frac{3\chi-1}{2} {\vec{n}~\vec{n}} \right ],
\label{Edfact}
\end{equation}
where ${\bf U}$ is the identity matrix, ${\vec{n}} := \frac{\vec{
f}}{f}$ and $\vec{f}$ the normalized energy flux defined as ${\vec f}
:= \frac{\vec{J}}{cu}$. In the limit of isotropic radiation
(Eddington limit), $\chi(f=0)=1/3$, while in the free streaming case
$\chi(f=1)=1$. A number of different expressions for the Eddington
factor have been introduced in the literature \cite{Lever} by
interpolating between this limiting 
cases. Some of them have been obtained from maximum entropy
principles. For instance, in \cite{Anile,Kremer} radiation under an
energy flux is studied by exploiting the entropy inequality,
i.e. by maximizing a generalized flux-dependent entropy under a set
of constraints, and an Eddington factor given by
\begin{equation}
\label{Edinf}
\chi = \frac{5}{3} - \frac{2}{3} \sqrt{4-3f^2}
\end{equation}
is obtained. The same result is recovered in \cite{QK} from a
information theoretical formalism, whereas different versions of this
formalism have been used in \cite{Todos,Fu} to obtain other variable
Eddington factors. 

Thus, we observe that a flux dependent pressure tensor with the form
\begin{equation}
\label{presion}
{\bf P}=a(u,J^2) {\bf U} + b(u,J^2) \vec{J} \vec{J},
\end{equation}
has been widely employed in radiative transfer. In order to obtain such a
dependence, which apparently departs from local equilibrium, some
authors \cite{QK,Kremer,Fu} have consider a flux-dependent
generalized entropy and Gibbs relation. In addition,
(\ref{presion}) also appears in the study of an
ultrarelativistic ideal gas under an energy flux by means of
Information Theory \cite{QK}.

However, in spite of what is claimed by the authors in \cite{QK,Anile,Kremer},
(\ref{Edinf}) can be obtained for the two simple cases of radiation and
an ultrarelativistic gas without abandoning the local equilibrium
hypothesis. The equations of state and entropies appearing in
\cite{QK,Anile,Kremer} may be recovered as well.

First of all, let us notice that these systems are submitted to an {\it energy
flux}, and not to a {\it heat flux},
because the condition of null global velocity has not been imposed.
Therefore, it is not difficult to show that the considered situation
corresponds to equilibrium (i.e., a purely advective energy
flux), in contradiction to what is assumed in
\cite{QK,Anile,Kremer}. In fact, due to the symmetry of the energy 
momentum tensor of a relativistic system (i.e. $T^{\mu \nu}=T^{\nu
\mu}$), the energy flux $\vec{J}$ verifies  
\begin{eqnarray}
\label{Flux}
\vec{J}=c^2 \vec{P}.
\end{eqnarray}
This property can also be obtained directly for a system of ideal
relativistic particles, with energy $\epsilon_i=\sqrt{m^2c^4+p_i^2c^2}$ and
velocity $\vec{v_i}=c \frac{\vec{p_i}}{\sqrt{p_i^2 + m^2 c^2}}$. The
total energy flux is  
\begin{equation}
\label{Micro1}
\vec{J}=\sum \epsilon_i \vec{v_i} =:\sum \vec{j_i},
\end{equation}
and introducing the expresions for $\epsilon_i$ and $\vec{v_i}$, it can be
readily verified that 
\begin{equation}
\label{Micro}
\vec{j_i}= \epsilon_i \vec{v_i} =c^2 \vec{p_i}.
\end{equation}
Therefore, equation (\ref{Flux}) holds for this system.

Following (\ref{Flux}), the equations of state for a system submitted
to an energy flux $\vec{J}$ (without any additional restriction on
the particle flow) must be the same as the equilibrium equations of
state of a moving system with momentum $\vec{P}$, which can be obtained
by simply performing a Lorentz boost to an equilibrium system at rest.
Therefore, the systems considered in \cite{QK,Anile,Kremer}
are nothing else but moving equilibrium systems.
The distribution functions of both cases can be obtained by the use of
Lorentz transformations as follows. The rest frame ($K_0$)
equilibrium distribution function can be written as:
\begin{equation}
\label{f0}
f=\frac{g}{e^{\alpha_0+\beta_0 \epsilon_0}+a},
\end{equation}
where $g$ is the degeneracy, $\alpha_0=-\beta_0\mu_0$ and
$a=-1$ for bosons, $a=1$ for fermions, $a=0$ for particles
obeying Boltzmann's statistics and $a=-1, \mu_0=0$ for photons and
phonons. We consider the cases of radiation 
and a classical ideal ultrarelativistic gas, so $\epsilon_0=p_0c$.
An observer at rest in a frame $K$ moving with momentum $-\vec{P}$ and
velocity $-\vec{V}$ with respect to the $K_0$ frame measures an
energy $\epsilon=pc$ for a particle with momentum $p$ (and velocity
$\vec{c}$) that verifies: 
\begin{equation}
\label{Lorentz}
\epsilon_0=\gamma\left(\epsilon-\vec{V}\vec{p}\right).
\end{equation}
Substitution of (\ref{Lorentz}) in (\ref{f0}) gives
\begin{equation}
\label{f1}
f=\frac{g}{e^{\alpha+\beta \epsilon+\vec{I}\vec{p}c^2}+a},
\end{equation}
where we have defined $\beta:=\gamma \beta_0$ and
$\vec{I}:=-\beta\vec{V}/c^2$. Note that 
$(\beta,\vec{I}c)$ is the so-called coldness 4-vector.
If we now use (\ref{Micro}) we obtain
\begin{equation}
\label{f}
f=\frac{g}{e^{\alpha+\beta \epsilon+\vec{I}\vec{j}}+a}.
\end{equation}
Now, we can recover the distribution function used in
\cite{QK} for radiation by simply setting $a=-1$,
$\alpha=0$, $\epsilon=pc$, $\vec{j}=\epsilon \vec{c}$, $g=2$:
\begin{equation}
\label{frad}
f=\frac{2}{e^{\beta pc+\vec{I}\vec{c}pc}-1},
\end{equation}
whereas for the classical ultrarelativistic gas we obtain the
distribution function proposed in \cite{QK} setting $a=0$.
Once this distribution function is fully justified, the whole
procedure in \cite{QK} holds.
Thus, the results obtained in \cite{QK,Anile,Kremer} are recovered and, in
particular, defining the pressure tensor as the mean value of the operator
\begin{equation}
\hat{P}_{\alpha \beta} := V^{-1} \sum_{i=1}^{N} p_i^{\alpha} v_i^{\beta},
\end{equation}
and using (\ref{Edfact}), (\ref{Edinf}) is obtained.
However, the physical interpretation given by this derivation is
completely different to that given in \cite{QK,Anile,Kremer}.

Clearly, the distribution of particles is anisotropic in the frame $K$
due to the additional vectorial constraint (the global momentum
$\vec{P}$). The distribution function (\ref{frad}) allows the study of
anisotropic equilibrium radiation (being the anisotropy due to the
relative motion) but not the study of nonequilibrium situations. We
propose the following heuristic argument to understand the physical
situation. Although in \cite{QK,Anile,Kremer} different methods were used
to arrive to the equations of state of nonequilibrium radiation
submitted to an energy flux, the authors never imposed the constraint
of no global motion of the system. Therefore, when they made use of the
condition of maximum entropy, they found an equilibrium moving system
because equilibrium situations have the maximum entropy and the
moving system verifies the imposed constraint of non zero energy
flux. 

Let us remark another interesting feature of (\ref{frad}) related to
the physical meaning of temperature in this moving system.
The distribution function (\ref{frad}) can also be viewed as a Planck
distribution with an effective $\beta_{ef}$ given by
\begin{equation}
\beta_{ef} := \beta + I c \cos \theta = \beta (1 - \frac{V}{c} \cos \theta).
\end{equation}
This expression is used, for instance, in cosmology in the study of
the Cosmic Microwave Backgroung Radiation (CMBR) in order to take
into account the relative movement between the Earth and the
reference frame defined by the CMBR. 
By averaging over the angular dependence with the distribution
function (\ref{frad}), it is obtained that
\begin{equation}
< \beta_{ef} > = \beta \left (1 - \frac{I^2}{\beta^2} \right) =
\frac{\beta}{\gamma^2} = \frac{\beta_0}{\gamma},
\end{equation}
so it is possible to define an effective mean temperature given by
\begin{equation}
T_{ef} := \frac{1}{k_B \beta_{ef}} = \gamma T_0,
\end{equation}
where $T_0:=\frac{1}{k_B \beta_0}$.
Therefore, $T_{ef}$ is found to simply be the Lorentz transformation
of $T_0$, according to Ott's transformation law \cite{Neuge}. This
gives a simple interpretation for Ott's temperature, whose physical
bases were controverted along the sixties \cite{Yuen}. 

In \cite{Essex}, it has been argued that in some situations it is not
possible to apply the methods of nonequilibrium thermodynamics for radiation.
This is the case, for example, for two planar
surfaces fixed at different temperatures $T_1$ and $T_2$ which exchange energy
through a radiation field. Photons travelling in one direction are
characterized by $T_1$, and the ones travelling in the opposite
direction by $T_2$, so it is not possible to assign a single
temperature to radiation. Based on these arguments, recently
\cite{Critica} has criticized the thermodynamical methods used in
\cite{Anile,Kremer} in their analysis of anisotropic radiation.
Hovewer, we have proved that the system considered in
\cite{Anile,Kremer} is, in fact, an equilibrium moving system and
therefore the criticism does not hold in this case. Let
us remark that in the system considered by Essex \cite{Essex} the
distribution function is characterized by a double peak, whose
relative heights depend on direction, while ours
has a single peak whose position varies with direction. Thus, it is
possible to define an angle-dependent effective temperature.

In addition, in \cite{Critica} the validity of expressions like
(\ref{presion}) for the pressure tensor has been questioned, both for
gases and for radiation. We have seen that in the cases of equilibrium
moving radiation or an ultrarelativistic gas, the pressure tensor adopts
an anisotropic form due to the presence of an additional vectorial
constraint (i.e. $\vec{J}$). We think that these simple problems can
serve as a guide to more complicated nonequilibrium situations.
Therefore, it seems a plausible possibility that for a
nonequilibrium system submitted to an energy flux and zero mass flow,
the pressure tensor also depend on the energy flux, like in
equilibrium. If that is the case, the dependence 
must have the form in (\ref{Edfact}) because, from a purely algebraic
point of view, the most general tensor that may be built up in
presence of a vector $\vec{J}$ must have the form $a(J^2){\bf
U}+b(J^2)\vec{J}\vec{J}$ and according to the definition for the
pressure tensor $tr {\bf P} = u$. However, this question remains
open, and such a form for the pressure tensor 
is not free of difficulties, as pointed out in \cite{Critica}.
Taking into account these criticisms, and the fact that expressions
of the form (\ref{presion}) are widely used in radiation
hydrodynamics, the convenience of finding a consistent thermodynamic
scenario for these systems arises.
We should also note that some variable Eddington factors $\chi$ have been
proposed \cite{Todos} using maximum entropy principles without
a careful interpretation of the generalized flux-dependent
entropies that naturally appear in the formalism.

A plausible framework to understand these nonequilibrium
flux-dependent entropies appearing in radiation transfer may be
Extended Irreversible Thermodynamics (EIT) \cite{Ext}.
According to EIT, both temperature and thermodynamic 
pressure should be modified in nonequilibrium situations, if a
generalized flux-dependent entropy function is considered. Up to
second order in the fluxes, one has
\begin{equation}
s(u,v,\vec{J})=s_{eq} (u,v) + \alpha(u,v) \vec{J}\cdot \vec{J}
\end{equation}
and if pressure and temperature are defined, as usual, by the
derivatives of the entropy function, one can easily obtain
flux-dependent equations of state:
\begin{equation} 
\frac{1}{\theta}= \frac{1}{T} + \frac{\partial \alpha}{\partial u}
\vec{J}\cdot \vec{J}, 
\end{equation}
\begin{equation}
\frac{\pi}{\theta}= \frac{p}{T} + \frac{\partial \alpha}{\partial v}
\vec{J}\cdot \vec{J},
\end{equation}
where $T$ is the kinetic or local-equilibrium temperature and $p$ the
local-equilibrium pressure and $\theta$ and $\pi$ their generalized
flux-dependent counterparts. In addition, the resulting pressure
tensor was supposed in \cite{QK} to adopt the form: 
\begin{equation}
\bf{P}= \pi {\bf U} + \psi \vec{J} \vec{J},
\end{equation}
where $\psi$ is determined by the requirement that $tr {\bf P}= u$.

\section*{Acknowledgements}
We thank Prof. D. Jou and Prof. Casas-V\'azquez from the
Autonomous University of Barcelona for their suggestions.\par
The authors are supported by doctoral scholarships from the
Programa de formaci\'o d'investigadors of
the Generalitat de Catalunya under grants FI/94-2.009 (R.D.) and
FI/96-2.683 (J.F.).  We also
acknowledge partial financement support from the Direcci\'on General
de Investigaci\'on of the Spanish Ministry of Education and Science
(grant PB94-0718) and the European Union under grant ERBCHRXCT 920007.

\end{document}